\documentclass[showkeys,nofootinbib,prd]{revtex4}

\usepackage{amsfonts}
\usepackage{amssymb}
\usepackage{natbib}
\usepackage[latin1]{inputenc}
\usepackage[babel]{csquotes}
\usepackage{graphicx}
\usepackage[english]{babel}
\usepackage{multirow}

\usepackage[T1]{fontenc}
\usepackage[centertags]{amsmath}

\usepackage{mathrsfs}
\usepackage{amsmath}
\usepackage{amsthm}
\usepackage{amssymb}
\usepackage{mathtools}
\usepackage{wasysym}
\usepackage{subfigure}
\usepackage{hyperref}

\begin{document}

\title{Electromagnetic emission from a black-to-white hole transition - Photospheric emission}

\author{Mattia Villani}
\affiliation{University of Urbino Carlo Bo, Department of Pure and Applied Sciences (DiSPeA), Via Santa Chiara, 27, Urbino (PU), 61029, Italy}
\email{mattia.villani@uniurb.it}

\begin{abstract}
We calculate the gamma ray emission from a black-to-white hole transition. We put forward a model for the prompt emission phase, based on the photospheric emission model with the underlying idea that the number of photons and leptons emitted by the white hole will be larger than the baryon content. We calculate the spectrum of the emitted radiation and give an estimate of the total energy emitted. {Finally, we estimate the cosmological density of lunar mass primordial black holes.}
\end{abstract}


\maketitle

\section{Introduction}
The possibility that black holes might end their life as white holes was first suggested in \citet{planck1} in the context of Loop Quantum Gravity, following the results obtained in Loop Quantum Cosmology by \citet{lqc} \citep[see also][for example]{bojowald}. The main idea is that the dying star keeps collapsing inside the event horizon until it reaches densities comparable to the Planck density: at this point (yet unknown) quantum effects revert the collapse causing the transition from a black hole to a white hole, i.e. they trigger an explosion. Due to the huge time dilation caused by the large gravitational field around the star, an external observer sees the collapsing star as a black hole for millions of years, while in the star reference frame, this whole process is very fast. In \citet{planck5} it was proved that in the observer reference frame the lifetime of the black hole is proportional to the square of the initial mass of the hole and that lunar mass objects ($\approx 10^{26}$ g) have a large probability to explode in the current period of the Universe history.

The matter ejected from the white hole will be hot and it will have relativistic speeds, thus there are the conditions for the formation of a fireball. It is therefore possible that a black-to-white hole transition might be observed in the  gamma-ray part of the electromagnetic spectrum. However, due to the lack of a large background magnetic field, it is likely that no jet will form, but, instead, a spherical shell of matter will expand at relativistic speeds from the initial position of the event horizon into the interstellar medium causing the formation of a shock. The possible increase of the interstellar magnetic field due to Weibel instability and its effect on the energy emission will be studied in a forthcoming paper.

We do not know the state of matter in the very first moments after the explosion, because most likely a theory of Quantum Gravity is needed; however, after some time, normal matter will form in the form of leptons and quarks. Quarks will quickly hadronize, forming mesons and baryons, however, they will quickly decay leaving behind only protons, neutrons, and leptons, mostly electrons and muons; neutral pions will decay in pairs of photons. Muons and tau particles will also decay leaving only electrons and positrons. In the present paper, we shall neglect the contribution of muons, which have a relatively long lifetime and could affect the evolution of the system; we focus only on electrons, positrons, and photons. We shall also neglect neutrinos which move into interstellar space undisturbed, carrying away energy.

The matter ejected from the white hole will likely be neutral\footnote{Any charge present in the initial black hole will quickly be neutralized by infalling matter, see however \citet{ch2,ch1,ch3}.}, thus:
\begin{equation}
n_{-}-n_{+} = n_p
\end{equation}
where $n_{-}$ and $n_{+}$ are the electron and positron density, respectively, $n_p$ is the proton density. Due to their smaller mass (and also because of the chain of decay of higher mass particles), the number of electrons and positrons will be much larger than that of other particles, thus the expanding shell will be dominated by electron--positron pairs and by photons. It is also likely that there will be neutrons with density $n_n \approx n_p$, but still $n_p+n_n \ll n_{-}+n_{+}$.

The fact that the fireball in a white hole will be radiation dominated with a low baryon load makes the photospheric emission process plausible, see \citet{pacz,good,piran,levinson,alamaa}, for example. In this process, photons are continuously produced by $e^\pm$ pair annihilation and Compton process, while their number will be constantly depleted at high energies by pair production. As a result, the radiation emitted once the fireball becomes optically thin will be non-thermal.

In this paper, we present a simple model in which we self-consistently take into account the various processes that produce and destroy photons and electron/positron pairs, such as pair annihilation and pair production and Compton process. We also take into account the adiabatic cooling of leptons as the shell of matter expands into the interstellar medium.

This paper is organized as follows: in Section \ref{sec:model} we describe our model for the dynamics of the evolution of electron/positron and photon densities; in Section \ref{eq:results} we describe how we solve our equations and our results; {in Section \ref{sec:PBH}, we compare our results with data from Integral and COMPTEL, in order to derive an upper limit to the cosmological density of primordial black holes that have gone through a black-to-white transition; }in Section \ref{sec:conclusion} we conclude our exposition.

\section{Model for photons and leptons dynamics}
\label{sec:model}

For photons, we consider the transfer equation \citep{hsieh,levinson2}:
\begin{equation}
k^\mu\, \dfrac{\partial n}{\partial x^\mu} = \dot{n}_{pair}+\dot{n}_C^++\dot{n}_C^--\dot{n}_{a/pair}
\end{equation}
where $\dot{n}_{pair}$ is the rate of photon production by pair annihilation, $\dot{n}_C$ is the rate of Compton scattering and the last term $\dot{n}_{a/pair}$ is the absorption rate of the photons by pair production. The 4-vector $k^\mu$ is given by:
\begin{align}
k^\mu &=\nu_0(1,\vec{l}_0),\\\label{eq:q}
\nu_0 &=\gamma\, \nu\,(1-\vec{v}\cdot\vec{l}),\\\label{eq:c}
\vec{l}_0 &= \dfrac{\nu}{\nu_0}\left[ \vec{l} + \left( \dfrac{\gamma-1}{v^2}\, (\vec{v}\cdot\vec{l})-\gamma \right)\, \vec{v} \right]
\end{align}
where $\vec{v}$ is the 3-velocity of the matter, $\gamma$ is its Lorentz factor, $\vec{l}$ (with $\vec{l}\cdot\vec{l}=1$) is the photon direction vector and $\nu$ is the photon frequency.

The positrons and electron densities, $n_{\pm}$, are calculated using a continuity equation as follows:
\begin{equation}
\dot{n}_{\pm}+ \vec{\nabla}\cdot(n_{\pm}\vec{u}_\pm)=-\dot{n}_{A}+\dot{n}_{pp}+\dot{n}_{CC},
\end{equation}
where $\dot{n}_{A}$ is the absorption of $e^{\pm}$ pairs due to annihilation, $\dot{n}_{pp}$ is the pair production term, $\dot{n}_{CC}$ is the Compton cooling term and $u_\pm^\mu$ is the 4-velocity of the particle, given by:
\begin{equation}
u_\pm^\mu = \gamma_\pm\,(1,\vec{u}_\pm), \qquad \gamma_\pm=\dfrac{1}{\sqrt{1-u_\pm^2}}.
\end{equation}

If we assume spherical symmetry and that particles propagate radially, we have
\begin{equation}
\vec{\nabla}\cdot(n_{\pm}\vec{u}_\pm) = \left.\dfrac{1}{r^2}\, \sqrt{1-\dfrac{1}{\gamma_\pm^2}}\,\partial_r\left( r^2\,n_{\pm} \right)\right|_{r=R},
\end{equation}
where $R$ is the radius of the shell. Moreover, in this case, the vector $\vec{l}$ in equations \eqref{eq:q} and \eqref{eq:c} is purely radial.

We now describe the various terms present in the above equations.

\subsection{Production of photons by annihilation of pairs}
Following \cite{svensson82}:
\begin{equation}
\dot{n}_{pair} = \int d\gamma_{+}\int d\gamma_{-} \, \left[n_+(\gamma_+)n_-(\gamma_-)\,\overline{v\, \dfrac{d\sigma}{dk}}\right],
\end{equation}
where
\begin{equation}
\overline{v\, \dfrac{d\sigma}{dk}}= \dfrac{\pi\,r_0^2}{\beta_+\beta_-\gamma_+^2\gamma_-^2}\, \left[ \sqrt{(\gamma_++\gamma_-)^2+\gamma_{cm}^2} + H_+ + H_- \right]_{\gamma_L}^{\gamma_U},
\end{equation}
with
\begin{equation}
H_\pm=\left( 2+\dfrac{1-\gamma*^2}{c_\pm} \right)\,I_\pm\, \left[ \dfrac{1}{\gamma_{cm}}-\dfrac{\gamma_{cm}}{2\gamma*^2}\,(2c_\pm-d_\pm) \right]\, \dfrac{1}{u_\pm}+\dfrac{\gamma_{cm}}{c_\pm}\,u_\pm,
\end{equation}
\begin{equation}
I_\pm=\left\{ \begin{array}{lr}
\dfrac{1}{\sqrt{c_\pm}}\, \ln\Big[ \gamma_{cm}\, \sqrt{c_\pm}+u_\pm \Big] & c_\pm>0\\
\dfrac{1}{\sqrt{-c_\pm}}\, \arcsin\left[ \dfrac{\gamma_{cm}}{\gamma*}\, \sqrt{-c_\pm} \right] & c_\pm<0
\end{array} \right.,
\end{equation}
while for $c_\pm=0$:
\begin{equation}
H_\pm = \left( \dfrac{2}{3}\gamma_{cm}^3+2\gamma_{cm}+\dfrac{1}{\gamma_{cm}} \right)\, \dfrac{1}{\gamma*} + \dfrac{1}{2} \left( \dfrac{2}{3}\, \gamma_{cm}^3-d_\pm \gamma_{cm} \right)\, \dfrac{1}{\gamma*^2}.
\end{equation}
The other functions are defined as follows:
\begin{align}
c_\pm &= (\gamma_\mp -x)^2-1,\\
u_\pm &= \sqrt{c_\pm\,\gamma_{cm}^2+\gamma*^2},\\
d_\pm &= \gamma_\mp\,(\gamma_++\gamma_-)\pm x\,(\gamma_++\gamma_-),\\
\gamma*^2 &= x\,(\gamma_++\gamma_--x),\\
\gamma_U &= \min[\gamma_{cm}^{\max},\gamma*],\\
\gamma_L &= \gamma_{cm}^{\min}.
\end{align}
$\gamma_{cm}$ is the energy in the center of mass, $\gamma_{cm}^{\max}$ is the maximum energy in the center of mass allowed by the kinematics, and analogously $\gamma_{cm}^{\min}$ is the minimum energy allowed. $x$ is the energy of the photon in units $m_ec^2$ where $m_e$ is the electron mass and $r_0$ is the classical radius of the electron. $\gamma_\pm$ are the Lorentz factor of positrons and electrons, respectively.

\subsection{Absorption of photons by pair production}

Following \cite{svensson83}, the absorption rate by pair production can be written as:
\begin{equation}
\dot{n}_{a/pair} = \dfrac{2n(x)}{x^2}\, \int^{\infty}_1 ds \left[ s\, \sigma(s)\, \int^{\infty}_{s/x} \dfrac{dy}{y^2}\,n(y) \right]
\end{equation}
where the cross section is given by:
\begin{equation}
\sigma(s) = \dfrac{\pi\,r_0^2}{s}\,\left[ \left( 2 + \dfrac{2}{s} - \dfrac{1}{s^2} \right) \, \text{arccosh}(\sqrt{s}) - \left( 1+\dfrac{1}{s} \right)\, \sqrt{1-\dfrac{1}{s}} \right].
\end{equation}

\subsection{Compton scattering}

For the Compton interaction, we consider \cite{blumenthal} and use the expression:
\begin{equation}
\dot{n}_{C}^\pm =  n_{\pm}(\gamma)\, \int dx^\prime \left( \dfrac{n(x^\prime)}{x^\prime} \, \int^{E_{max}}_{E_{min}} \sigma_{KN}(E,\gamma) dE \right)
\end{equation}
where
\begin{equation}
E_{min} = \dfrac{x^\prime}{\gamma}, \qquad E_{max} = \dfrac{4x^\prime\,\gamma}{1+4x^\prime\,\gamma}
\end{equation}
and
\begin{equation}
\sigma_{KN} = \dfrac{2\pi\, r_0^2}{\gamma}\, \left[ 2q\,\ln(q)+(1+2q)\,(1-q)+\dfrac{1}{2} \dfrac{(\Gamma q)^2}{1+\Gamma q}\,(1-q) \right]
\end{equation}
with
\begin{equation}
\Gamma=4x^\prime\, \gamma, \qquad q=\dfrac{E}{\Gamma\,(1-E)}, \qquad E<1 
\end{equation}

\subsection{Annihilation of pairs of leptons}
For the annihilation of positron/electron pairs, we have considered \citep{svensson82}
\begin{equation}
\dot{n}_{A} = n_\pm \int d\gamma_\mp n_{\mp}(\gamma_\mp)\, \overline{v\sigma}
\end{equation}
where
\begin{equation}
\overline{v\sigma} = \dfrac{\pi\, r_0^2}{\beta_+\gamma^2_+\beta_-\gamma^2_-}\, \Bigg[ \beta^3\,\gamma^2\, \ln\left( \dfrac{1+\beta}{1-\beta} \right) -2\gamma^2+\dfrac{3}{4} \, \left[\ln\left( \dfrac{1+\beta}{1-\beta} \right)\right]^2 \Bigg]_{\gamma_{min}}^{\gamma_{max}}
\end{equation}
\begin{align}
\beta&=\sqrt{1-\dfrac{1}{\gamma^2}}\\
\gamma_{min} &= \sqrt{\dfrac{1}{2} \, \Big[ 1+\gamma_+\gamma_--\gamma_+\gamma_-\beta_+\beta_- \Big]}\\
\gamma_{max} &= \sqrt{\dfrac{1}{2} \, \Big[ 1+\gamma_+\gamma_-+\gamma_+\gamma_-\beta_+\beta_- \Big]}.
\end{align}

\subsection{Pair production for leptons}

For pair production, we have considered \citep{GS}:
\begin{equation}
\dot{n}_{pp} = \int dx^\prime \dfrac{n(x^\prime)}{(x^\prime)^2} \, \int^{2x\,x^\prime}_1 ds\,s\, \sigma_{pp}(s) 
\end{equation}
with
\begin{equation}
\sigma_{pp} = \pi\, r_0^2\, (1-\beta^2)\, \left[ (3-\beta^4)\, \ln\left( \dfrac{1+\beta}{1-\beta} \right)\, 2\beta\,(2-\beta^2) \right]
\end{equation}
\begin{equation}
\beta=\sqrt{1-\dfrac{1}{s}}.
\end{equation}

\subsection{Cooling of electrons and positrons}
The cooling of electrons and positrons by Compton scattering can be modeled as \citep{lightman87}:
\begin{equation}
\dot{n}_{CC} = \dfrac{\partial}{\partial\gamma}\left[ \dot{\gamma}\, n_{e^\pm} \right]
\end{equation}
where
\begin{equation}
\dot{\gamma} = -c\,\sigma_T\,\left( \dfrac{4}{3}\gamma^2-1 \right) \, \int^{3/(4\gamma)}_0n(x)dx.
\end{equation}

In the last term, we also add an adiabatic cooling due to the expansion of the shell \citep{uhm,geng,geng2}:
\begin{equation}
\dot{\gamma} = -\dfrac{2}{3}\,\dfrac{\gamma}{R} \, \dot{R},
\end{equation}
where $R$ is the radius of the shell and $\dot{R}$ is its expansion velocity. We assume that:
\begin{equation}
\dot{R} = c\,\sqrt{1-\dfrac{1}{\gamma^2}}.
\end{equation}

\subsection{The use of Minkowski spacetime}
{In all of the above equations, we have implicitly assumed that the underlying spacetime is Minkowski. This is justified in our case, since the BH considered has a mass of $10^{26}$ g, and therefore a Schwarzschild radius of $10^{-4}$ m. The matter is ejected by the WH at a considerable faction of the speed of light; thus, it quickly reaches a point where all general relativistic effects, redshift included, are completely negligible. This will no longer be true once we consider larger BH. In this case, one should consider the metric of an expanding sphere of radiation, for example, the time reversed version of the metric considered in \cite{bojo} and consistently modify all the above equations, substituting partial derivatives with covariant derivatives; in this way all relativistic effects will be automatically included. This is left for a future work.}

\section{Method of solution and results}
\label{eq:results} 

\subsection{Method}
In order to solve the system of equations described in the above Section, we have employed the splitting scheme for partial differential equations (PDE) with source terms \citep[see][]{toro}. The scheme has the following form: given a 1-D PDE with sources $S$ of the form
\begin{equation}
\partial_t u+\partial_r f(u) = S(u)
\end{equation}
and initial condition (IC)
\begin{equation}
u(t^n,r)=u_n,
\end{equation}
we split it into a PDE and an ordinary differential equation (ODE) as follows:
\begin{align}\label{eq:pde}
\text{PDE:   }& \partial_t u + \partial_r = 0,\\
\text{IC:    }& u(t^n,r)=u_n,
\end{align}
which has solution $\overline{u}_{n+1}$ and
\begin{align}\label{eq:ode}
\text{ODE:   }& \dfrac{du}{dt} = S(u),\\
\text{IC:    }& u(t^n,r)=\overline{u}_{n+1},
\end{align}
with solution $u_{n+1}$, which becomes the IC for the PDE in the next time step. We treat the PDE \eqref{eq:pde} with a Godunov scheme \citep[see][]{toro} and we solve the ODE \eqref{eq:ode} with a second order Runge-Kutta method. The energy dependence is treated with the method of lines. We run our program until the optical depth reaches 1: at this point we assume that there is the emission of the radiation.

As initial conditions, we consider a thermal distribution for both photons and leptons, in particular, we consider a Bose-Einstein distribution for photons and a Fermi-Dirac distribution for leptons. We have several parameters to set: the baryon (proton plus neutron) fraction, the initial fraction of pions (necessary to set the initial photon density), and the temperature. We assume that the fraction of baryons is 1\% of the fraction of leptons. For the fraction of pions we consider three cases $n_{\pi}=\{5\%,10\%,20\%\}$ of the lepton density. We also consider two temperatures $k_BT=\{m_ec^2,2\,m_ec^2\}$. The initial density of the matter ejected for the white hole is huge, so we start our program when the matter has expanded and the density has reached $10^{14}$ g cm$^{-3}$, because this is the density of a neutron star and the physics is known. We consider an initial mass of $10^{22}$ kg (corresponding to the mass of the Moon), thus the matter will initially be compressed to a sphere with a radius of $0.1$ km.

\subsection{Results}

\begin{figure}
    \centering
    \includegraphics[width=0.85\linewidth]{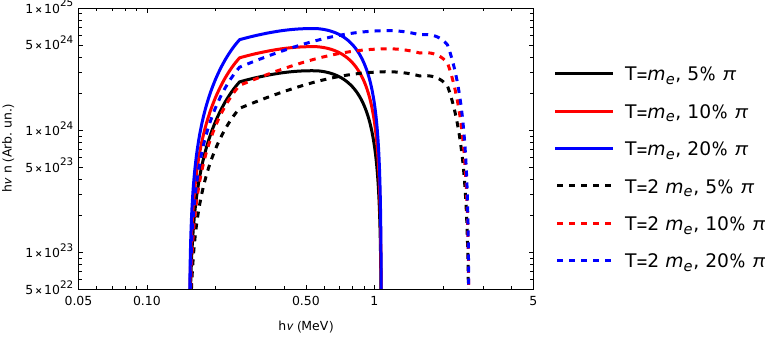}
    \caption{The spectra of the emitted photons. We report in black the case of 5\% of pions, in red the case of 10\% of pions, and in blue the case of 20\% of pions. Thick curves are for $k_BT=m_ec^2$, while dashed curves for the case $k_BT=2m_ec^2$.}
    \label{fig:spectra}
\end{figure}

The spectra of the emitted photons are reported in figure \ref{fig:spectra}. We notice that in all the considered cases, the spectra are non-thermal with a peak around 0.5 MeV for $k_BT=m_ec^2$ and around 1 MeV for $k_BT=2m_ec^2$; there are sharp cutoffs in the spectra around 1 MeV and 3 MeV for the two temperatures we have considered. We notice that increasing the pion fraction increases the value of the peaks: this is because the initial fraction of photons increases due to the decay $\pi_0\rightarrow 2\gamma.$ We find that the energy emitted, reported in table \ref{tab:ener} is always larger than $\times 10^{43}$ erg. 

\begin{table}[ht]
    \centering
    \begin{tabular}{c|cc}
        Fraction of pions & $k_BT=m_ec^2$ & $k_BT=2m_ec^2$ \\
        \hline
        5\%  & $5\times 10^{43}$ erg & $1.4\times 10^{44}$ erg\\
        10\% & $8\times 10^{43}$ erg & $2.2\times 10^{44}$ erg\\
        20\% & $3\times 10^{44}$ erg & $3\times 10^{44}$ erg
    \end{tabular}
    \caption{Total energy emitted from the burst.}
    \label{tab:ener}
\end{table}

\section{Estimate of the cosmological density of primordial black holes}
\label{sec:PBH}

\begin{figure}
    \centering
    \includegraphics[width=0.75\linewidth]{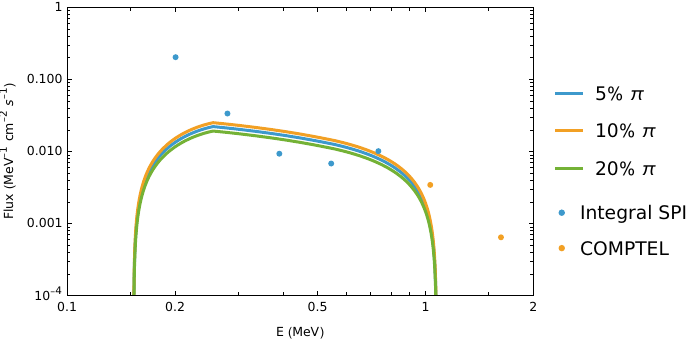}
    \caption{Comparison of the average flux of X- and $\gamma$-ray flux with the estimated flux due to the cosmological emission of radiation due to the black-to-white hole transition. Case of $T=m_e$.}
    \label{fig:PBH1}
\end{figure}

\begin{figure}
    \centering
    \includegraphics[width=0.75\linewidth]{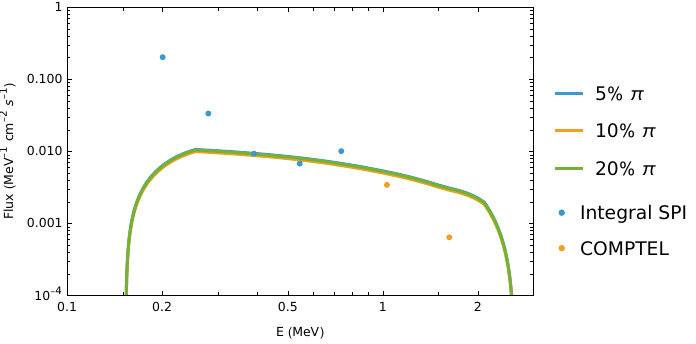}
    \caption{Comparison of the average flux of X- and $\gamma$-ray flux with the estimated flux due to the cosmological emission of radiation due to the black-to-white hole transition. Case of $T=2 m_e$.}
    \label{fig:PBH2}
\end{figure}

{In this section, we estimate the cosmological density of lunar mass primordial BH that have gone through the black-to-white hole transition. We compare the differential flux of X- and $\gamma$-rays measured by Integral SPI \cite{int1} and COMPTEL \cite{compt1} in the energy range $0.2-2$ MeV (see Figure \ref{fig:spectra}). We estimate the diffuse cosmological flux of radiation due to the black-to-white hole transition with the formula:}
\begin{equation}\label{eq:flux}
    \dfrac{d\Phi}{dEd\Omega}=\dfrac{1}{4\pi} \dfrac{\Gamma\,\Omega_{PBH}\rho_c}{M}\,\int_0^{10}dz\,\dfrac{dn}{dE}\,\dfrac{d\chi}{dz}
\end{equation}
{where $\Gamma=M^{-2}\,G^{-2}\,h^{1/2}\,c^{7/2}$ is the transition rate \cite{planck5}, $\rho_c$ is the critical density, $\Omega_{PBH}$ is the cosmological density of primordial BH we want to estimate, $M=10^{26}$ g is the initial BH mass; $dn/dE$ is the redshifted energy spectrum of emitted photons, and }
\begin{equation}
    \dfrac{d\chi}{dz}=\dfrac{c}{H_0}\,\dfrac{1}{\sqrt{\Omega_m\,(1+z)^3+\Omega_\Lambda}}
\end{equation}
{is the comoving distance. We assume $H_0=67.4$ km s$^{-1}$ Mpc$^{-1}$, $\Omega_m=0.315$ and $\Omega_\Lambda=0.685$ \cite{planck}. We fit equation \eqref{eq:flux} to Integral and COMPTEL data using least square method to find $\Omega_{PBH}$. The result of our analysis is given in Figures \ref{fig:PBH1} and \ref{fig:PBH2} and in Table \ref{tab:density}. From the figures, we notice that the case of $T=m_e$ seems to give the best fit to the data in the energy range $0.3-1$ MeV. From Table \ref{tab:density}, we notice that the cosmological density of primordial BH is larger ($\Omega_{PBH}\approx2.5\times10^{-8}$) for the case of 5\% of initial contribution of pions; the smallest density is obtained with $T=2m_e$ and 20\% of pions ($\Omega_{PBH}=0.90\times10^{-8}\pm10^{-10}$). We notice that this is an upper limit to the density and is very small, but not zero; thus the existence of these BH is not excluded, but they cannot constitute all dark matter.}

\begin{table}[htb]
    \centering
    \begin{tabular}{c|c||c|c}
         \multicolumn{2}{c||}{$T=m_e$}& \multicolumn{2}{c}{$T=2m_e$} \\
         \% of pions & $\Omega_{PBH}$ & \% of pions & $\Omega_{PBH}$\\
         \hline
         5  & $2.49\times 10^{-8}\pm10^{-10}$ & 5  & $1.98\times10^{-8}\pm10^{-10}$\\
         10 & $1.78\times 10^{-8}\pm10^{-10}$ & 10 & $1.20\times10^{-8}\pm10^{-10}$\\
         20 & $0.98\times 10^{-8}\pm10^{-10}$ & 20 & $0.90\times10^{-8}\pm10^{-10}$
    \end{tabular}
    \caption{Estimate of the cosmological density of lunar mass primordial BH.}
    \label{tab:density}
\end{table}

\section{Conclusions}
\label{sec:conclusion}
We have proposed a model for the study of the emission of electromagnetic radiation from a black-to-white hole transition based on the photospheric emission model. We have included in our model pair production, pair annihilation, Compton scattering of photons and Compton cooling and adiabatic cooling of leptons. Using as initial state a thermal distribution both for photons and leptons, we have solved our equations with the splitting scheme treating the energy dependence with the method of lines. We have evolved the system up to the moment in which the optical depth is below 1. We have found that the spectra of the emitted photons are non-thermal, with sharp cutoffs at 1 MeV or 3 MeV, depending on the initial temperature of the ejected matter. The total energy emitted is larger than $10^{43}$ erg for all the cases we have considered, however, we notice that a larger black hole would produce a larger burst of energy. As stated in the introduction, there will not be a jet, but the burst of radiation will be more or less isotropic. We notice that for the explosion of a white hole at cosmological distances, one should consider the cosmological redshift, which brings the peak of the emission close to the hard X-ray region of the electromagnetic spectrum. {We have also estimated the cosmological density of lunar mass primordial BH that have gone through a black-to-white hole transition, finding that it can go from $\approx10^{-8}$ to $\approx 2.5\times10^{-8}$, much smaller than the dark matter density, suggesting that the existence of primordial BH is not excluded, but they cannot constitute all dark matter.} In a forthcoming paper, we shall refine our work and include in our model the effects of synchrotron radiation produced by the enhancement of the interstellar magnetic field due to the Weibel instability.

\section*{Funding detail}
Funding: not applicable.



\end{document}